# Intelligent audit code generation from free text in the context of neurosurgery

**Sedigheh Khademi**
Faculty of Information Technology
Monash University
Victoria, Australia
Email: sedigh.khademi@gmail.com

**Pari Delir Haghighi**
Faculty of Information Technology
Monash University
Victoria, Australia
Email: pari.delirhaghighi@monash.edu

**Philip Lewis**
Department of Neurosurgery, Alfred Hospital
Monash Institute of Medical Engineering
Melbourne, Australia
Email: p.lewis@alfred.org.au

**Frada Burstein**
Centre for Organisational and Social Informatics
Faculty of Information Technology
Monash University
Victoria, Australia
Email: frada.burstein@sims.monash.edu.au,

**Christopher Palmer**
Christopher Palmer Limited
Email: chris.palmer.nz@gmail.com

## Abstract

Clinical auditing requires codified data for aggregation and analysis of patterns. However in the medical domain obtaining structured data can be difficult as the most natural, expressive and comprehensive way to record a clinical encounter is through natural language. The task of creating structured data from naturally expressed information is known as information extraction. Specialised areas of medicine use their own language and data structures; the translation process has unique challenges, and often requires a fresh approach. This research is devoted to creating a novel semi-automated method for generating codified auditing data from clinical notes recorded in a neurosurgical department in an Australian teaching hospital. The method encapsulates specialist knowledge in rules that instantaneously make precise decisions for the majority of the matches, followed up by dictionary-based matching of the remaining text.

# 1 INTRODUCTION

Auditing is a quality improvement process that seeks to improve patient care and outcomes through systematic review of care against explicit criteria (*Principles for Best Practice in Clinical Audit* 2002). The audit process requires codified data for aggregation and analysis of patterns. This data typically comes from electronic health records (EHRs) or some other unique codified system required by a speciality.

Despite the obvious advantage of codified records, they are often inaccurate, incomplete and not properly maintained (Kaplan 2007); and free-form text remains a key component of electronic health records. The use of free-form text and a parallel lack of use of standard terms in text-based electronic health records is extensive, and is a source of poor data quality and an inability to share data between systems, to construct decision support systems, and to make secondary uses of data (Price et al. 2013). Free-form text is inevitably idiosyncratic and also very often incomplete, and lacking any coding structure does not allow for aggregation and analytical comparison.

Notwithstanding the advantage of collecting structured data in clinical software, research indicates that clinicians value the narrative expressivity and workflow efficiency of entering free-form text (Rosenbloom et al. 2011). Systems designed to acquire structured data in real-time often have unnatural, inflexible, or inefficient user interfaces that place too much of a burden on busy clinicians, therefore it may be better to leverage computing technologies to extract codified data from free-form clinical notes using post-hoc text processing (Ash et al. 2004).

Researchers and developers of clinical information systems have used a range of technologies to try to achieve complete and accurate coded clinical data using post-hoc text processing. Some have used natural language processing (Long 2005)(Meystre and Haug 2006)(Long 2005), others have used data mining and machine learning techniques (Pakhomov et al. 2006)(Wright et al. 2010). Rosenbloom et al (Rosenbloom et al. 2011) suggest that we need to develop hybrid systems that combine structured entry with later text-processing.

This research aims to investigate techniques for processing the free text entered in a neurosurgical department of a major trauma hospital, in order to improve the codification of neurosurgical text data for the purposes of auditing. The research focuses on what kind of system is required to integrate seamlessly into the auditing process, efficiently and effectively meeting highly specific requirements, delivering immediately useful audit data from free text. The system design aims to be adaptable for integration into the upstream diagnostic code and note entry system.

# 2 RESEARCH CONTEXT

The Neurosurgical Department has an application that is used to describe the diagnoses and procedures performed for each patient passing through the department. This is independent of the main electronic records system of the hospital, and the data collected and terms used are highly specific to neurological conditions and neurosurgical procedures. The medical terminology used is closely aligned to that published by the Royal College of Surgeons, in order to support regular reporting and surgical activity tracking.

Our dataset spans ten years from mid-2003 to the end of 2014; the diagnostic data is contained in a table of 24,437 records, and the procedural data table has 11,721 records. The diagnostic codes have six levels of hierarchy with 6 root nodes and a total of 213 diagnoses. The 6 root nodes include such categories as "Spinal" "Cranial" or "Complications." Procedural codes have only two levels of hierarchy but 12 root nodes with a total of 134 procedures. Each diagnosis is associated with an audit category, which is very often a higher level classification that encompasses a number of diagnoses, but is sometimes as specific as the diagnosis; examples are given in table 1.

| Diagnosis Code | Diagnosis Value | Audit Category |
|---|---|---|
| 218-224 | Cranial>Trauma | CRANIAL:TRAUMA |
| 218-224-309 | Cranial>Trauma>Osseous Injury | CRANIAL:TRAUMA:SKULL FRACTURE |
| 218-224-309-310 | Cranial>Trauma>Osseous Injury>Skull | CRANIAL:TRAUMA:SKULL FRACTURE |
| 218-224-309-310-314 | Cranial>Trauma>Osseous Injury>Skull>Non-displaced | CRANIAL:TRAUMA:SKULL FRACTURE |
| 218-224-309-310-315 | Cranial>Trauma>Osseous Injury>Skull>Depressed | CRANIAL:TRAUMA:SKULL FRACTURE |
| 218-224-309-310-315-316 | Cranial>Trauma>Osseous Injury>Skull>Depressed>Open | CRANIAL:TRAUMA:SKULL FRACTURE |
| 218-224-309-311 | Cranial>Trauma>Osseous Injury>Sinuses | CRANIAL:TRAUMA:SKULL FRACTURE |
| 218-224-309-312 | Cranial>Trauma>Osseous Injury>Base of Skull | CRANIAL:TRAUMA:SKULL FRACTURE |
| 218-224-309-350 | Cranial>Trauma>Osseous Injury>Facial | CRANIAL:TRAUMA:SKULL FRACTURE |
| 218-220-251 | Cranial>Neoplasia>Extrinsic | CRANIAL:NEOPLASIA |
| 218-220-251-243 | Cranial>Neoplasia>Extrinsic>Acoustic Neuroma | CRANIAL:NEOPLASIA |
| 218-220-251-242 | Cranial>Neoplasia>Extrinsic>Meningioma | CRANIAL:NEOPLASIA:MENINGIOMA |
| 218-220-251-244 | Cranial>Neoplasia>Extrinsic>Pituitary | CRANIAL:NEOPLASIA:PITUITARY |

*Table 1: Diagnosis Code and Audit Category structure*

Yearly auditing is performed using fine-grained data for selected common diagnoses, and coarse data for others. With this in mind, we have structured our data mapping algorithm around the pattern of analysis derived from previous yearly audits undertaken by the neurosurgery department. Thus some individual diagnoses map to a single, high-level category, whilst others map to individual categories.

The structure of a record in the application is one of a code and an accompanying note, with as many records per admission as is required to properly code all of the diagnoses and procedures, though typically only one record exists per admission. A note is not required against a code, though it is expected that a note will appropriately qualify and amplify the code picked. Table 2 is an example of an admission record which uses one of the diagnoses illustrated in table 1, admission and patient codes are de-identified.

| Admission Code | Date | Diagnosis | Notes |
|---|---|---|---|
| 3301458954811 | xx/xx/xxxx | Cranial>Trauma>Osseous Injury>Skull>Depressed>Open | Ped v car left frontal depressed fracture, GCS 3, ETOH |

*Table 2: Example of an admission record*

## 3 RELATED WORK

Creating coded data from free-form text is comprised of Information Extraction and Mapping tasks. Information Extraction (IE) typically requires a pre-processing stage to clean the text and prepare it for processing, which then utilizes various techniques to categorize the text into entities of interest: a task called Named Entity Recognition (NER) (Meystre et al. 2008). Those named entities (NE) can then be mapped to their corresponding concepts in standard terminologies and used for creating codes.

Pre-processing includes document structure analysis, spell checking, sentence splitting and word tokenization, part-of-speech tagging, word sense disambiguation, and parsing to identify words of interest (Demner-Fushman et al. 2009).

In the clinical environment named entities are typically categorized as symptoms, investigations, test results, diagnoses, prognoses, drugs, treatments and procedures, and outcomes of treatments and procedures (Wang and Patrick 2009).

Various paths to NER can be found in the clinical literature - they generally use one or a combination (hybrid) of three approaches: rule-based, dictionary-based, and machine learning-based approaches (Krauthammer and Nenadic 2004). These can all be incorporated into Natural Language Processing (NLP) systems - which is a popular approach used by software tools that have been developed to deal with the codification of free-form text (Agah 2013).

Research shows that these various systems perform relative to the complexity of the task and desired outcome - for instance while an NLP based system may deal well with descriptive language it needs additional components to deal satisfactorily with structured items such as laboratory test results. Therefore it remains difficult to reach conclusions about the most effective tool, and there is no general uptake of these tools in clinical practice (Mary H Stanfill et al. 2010).

## 3.1 Information Extraction

Information Extraction (IE) is a process of retrieving specific targeted information from texts or speech and presenting them as fixed-format and unambiguous data. Like Information Retrieval (IR) it analyses the text for patterns using natural language techniques, but whereas IR will just return a series of documents matching a query, IE will return specific data from the documents (Cunningham 2005).

Unlike "full text understanding" which attempts to represent all of the information in a text, Information Extraction is limited in its output: with IE we specify in advance what is required – the semantic range of the output, the relations we want to represent, and the allowable data for each component of the output (Grishman 1997). Depending on the complexity of the domain and the inputs required by further processing steps, it may be sufficient to obtain just the resolved named entities from text.

## 3.2 Information Extraction in the Clinical domain

When applying information extraction techniques to clinical notes there are a number of unique challenges to overcome, as clinical texts have specific characteristics.

Sentence boundary detection is made more difficult because of the lack of natural sentence structures. Abbreviations, terse sentence structures with a lack of punctuation or narrative flow, and the use of medical titles, lists, frequent non-alphanumeric characters, drug doses etc. all confound sentence structure detection. Single characters or combinations of very few letters may in fact be significant abbreviations, whereas others may not be – in any case in an ordinary NLP task they would be discarded. Some of these may be placed in unusual places compared to standard texts – such as "?" before a word to indicate uncertainty of the following diagnosis – in ordinary language "?" would indicate the end of a sentence.

Medical texts are frequently qualified with modifiers suggesting uncertainty or possibility - using phrases like "indicative of", "possible", "less likely", and with negations. Some of these may be abbreviated – e.g. "NAD" meaning "no abnormality detected", which is often used in relation to a preceding text.

Entity recognition is made more difficult by variations in word order and spelling (e.g. "C7 right superior articular facet #" and "#R C7 sup art facet"), the use of synonymous and polysemous words, and abbreviations and acronyms - which can be specific to a particular institution (Xu et al. 2009), or area of practice, or even to a practitioner. Abbreviations themselves are often polysemous and at least a third of them have more than one sense (Liu et al. 2001).

Compounding these problems, there are often misspellings, in part due to the large number of second-language learners of English working in the health professions, but also to there being few built-in spell checkers in the medical data entry systems.

If it is possible to define domain and problem specific terms of interest, such as identifying diagnoses pertinent to neurosurgery, then algorithms and dictionaries can be purpose built for the task of finding terms of interest, and other features of the text can be given lesser priority - though still assessed for patterns that can contribute to the extracted information.

## 3.3 Information Extraction methods

### 3.3.1 Pattern matching

Pattern matching techniques (McNaught and Black 2006) such as regular expressions and the use of dictionaries of synonyms can perform well, especially if the text either conforms to a limited set of expressions such as in a report of test results, or if the text is very telegraphic and uses abbreviations repeatedly.

Pattern matching techniques have been successfully used to solve clinical information extraction problems (Turchin et al. 2006); (Friedlin and McDonald 2006) (Turchin et al. 2006). (Napolitano et al. 2010) describe a good example of the suitability of pattern matching using regular expressions for extracting information from structured text. Looking for variants of descriptions of a Gleason score, they were able to achieve 98% recall and almost 100% precision.

Text suitable for the application of pattern matching techniques will be abbreviated, in the form of lists or terse language which repetitively uses abbreviated terms, and as such lacks information that would make it appropriate for a deeper language analysis. If however the text is more natural or has contextual clues it becomes amenable to more sophisticated sentence structure analysis. Using natural language analysis the text can be processed to get a complete knowledge of how phrases are constructed and combined (syntactic information), and how they should be interpreted (semantic understanding).

### 3.3.2 Syntactic/Semantic

The majority of the systems used in clinical text processing combine syntactic and semantic analysis. Frequently cited are Medlee, MetaMap and cTAKES - which is cited more recently. cTAKES (Clinical Text Analysis and Knowledge Extraction System), (Savova et al. 2010) is an open-source NLP system developed at the Mayo Clinic using Apache UIMA and Apache Open NLP natural language processing toolkits. It processes clinical notes using a variety of rule-based and machine learning methods to identify clinical named entities by using UMLS or other dictionaries - classifying the entities as medications, diseases/disorders, signs/symptoms, anatomical sites and procedures.

### 3.3.3 Machine Learning

Almost all of the recent systems incorporate some machine learning methods, as a processing step or as the basis for the system. Many of these systems have been developed in response to "challenges" proposed by medical institutions, in order to encourage research. In 2007 the Computational Medicine Centre (CMC) of the University of Cincinnati created the International Challenge on Classifying Clinical Free Text Using Natural Language Processing - a shared task challenge to develop systems that could process free-text radiological reports and assign one or two labels from a set of 45 ICD-9-CM codes (ICD-9-CM: The International Classification of Diseases, 9th Revision, Clinical Modification). (Pestian et al. 2007) describe the majority of systems as using a medically based feature engineering component combined with machine learning methods. (Farkas and Szarvas 2008) placed first in the challenge with a hybrid hand-crafted and machine learning system. They suggested that incorporating machine learning components would be optimal in a system that needed to deal with thousands of codes.

In another international challenge, the i2b2 (Informatics for Integrating Biology and Bedside) 2010 Relations Challenge, the general trend was to design systems that were ensembles of complementary approaches, with rule-based systems providing input to a machine learning component, or rule-based post processing of machine learning output (Uzuner et al. 2011, p. 2).

### 3.4 Evaluation

We have tested natural language text processing systems that utilize standardized coding ontologies against the free text found in the neurosurgical clinical database, and have found that they are not a good fit. The neurosurgical department's free-text data is highly abbreviated and contains discipline-specific jargon, with the result that many of the terms used are not identified by standard systems, or are misidentified. Furthermore the department uses a simplified in-house coding ontology, and it is inefficient to translate to this from a standard ontology. Most importantly the requirement to extract just the audit and diagnostic data from the free text suits a targeted information extraction approach, rather than a more general processing of all of the text.

The kind of system required for the neurosurgical department audit code generation could be obtained by extensively fitting an existing generalized system, or by creating a purpose-built system; our conclusion was that the strongest case was for the latter. That is, to create a solution that is lightweight, responsive and easy to interact with, delivering precise matching from the parts of the text that matter, using pattern matching and machine learning components, and capable of being quickly tuned and improved.

## 4   RESEARCH DESIGN

We conducted this research according to Design Science principles, as described by (Hevner et al. 2004) and by (Peffers et al. 2007). Design Science is described as the process of creating and evaluating IT artefacts which are created to address previously unsolved organizational problems. An artefact may be described as a construct, a model, a method, or an instantiation; we have developed a method.

### 4.1   Method Design

The initial objective of our method is to enable comprehensive and accurate extraction of codified data suitable for the audit process from the neurosurgical department's records. The ultimate objective is for our method to be utilised at initial data entry to suggest appropriate diagnostic codes, which are more detailed and varied than the auditing codes. Our artefact is a computer programme designed to generate audit compatible classification suggestions based on the notes attached to a record, to enhance the process of reviewing the notes by a system expert at the time of auditing. The general architecture of the programme is shown in Figure 1.

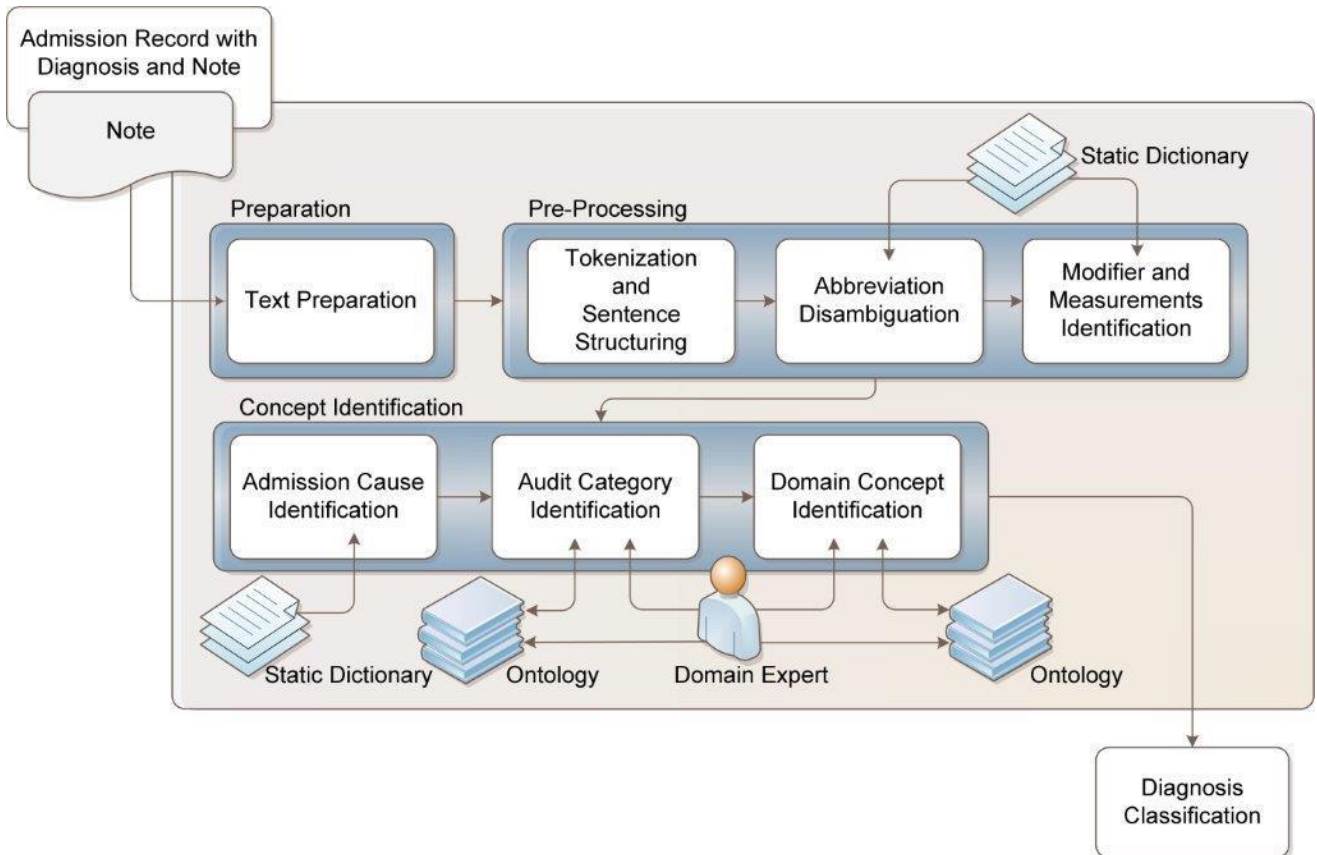

*Figure 1: Architecture of the audit code classification method*

To do this we need to first regularize the structure and language of the note, then break it into its constituent parts, and finally to identify those parts as belonging to certain concepts - so that we can find which of them are useful for deriving audit codes and diagnoses codes. Eventually we will make the diagnoses from a classification process utilizing the identified concepts. The process of determining our word meanings and concept identification is broken into three stages – Preparation, Pre-processing, and Concept Identification.

The audit code classification method consists of three consecutive stages, with the output of a stage providing the input for the next stage: Preparation of the text, followed by pre-processing, then by concept identification. Preparation ensures that the text is ready for the various steps in pre-processing, especially tokenization which requires that words are properly separated from one another. Pre-processing consists of all the steps that break the text into words and sentences, and there is also some identification of words performed, based on static dictionaries of words. Concept Identification is where useful words and phrases are identified – the outcome of this is that words and phrases are tagged for audit coding purposes. Each step here utilizes either a static dictionary of terms, or an ontology – meaning a dictionary of medical terms that is continuously being refined with feedback from a domain expert after analysis of the outputs of concept identification. The output of the audit code classification method is identification of audit categories, plus data suitable for the separate step of diagnosis coding, which can be developed after completion of this method.

### 4.1.1 Preparation

The Preparation step includes: Setting up the environment including dictionaries, fixing text boundaries prior to tokenization, abbreviation disambiguation, spelling correction and regularization of key words.

### 4.1.2 Pre-processing

Pre-Processing step includes: tokenization and sentence structuring, modifier and measurement identification, and abbreviation disambiguation.

### 4.1.3 Concept Identification

Concept Identification step includes: Admission Cause identification, Audit Category identification and Domain Concept identification.

After the Pre-processing we have words collected into groups within sentences, ready for further processing. Some of the words have already been identified, those which remain are therefore potentially meaningful for labelling for audit categories and diagnosis, and the labelling is the task of the Concept Identification stage.

Admission Cause identification is performed first, as there is value in identifying as much as possible of the note that does not directly relate to the identification of diagnosis, so that there are fewer remaining words for the Audit Category identification stage to analyse. It is also possible that an understanding of the cause of admission could help in identifying an Audit Category. We employ two overlapping static dictionaries of terms that allow these sorts of phrases to be identified.

Audit Category identification is a routine that examines the words one at a time, to see if they can be ascribed to an audit category, based on rules established by a system expert. If necessary the word being dealt with will be compared to other words in its containing sentence or even to all of the key words found in the Note. A summary of the initial audit categories is shown in Table 3.

| Concept (Cranial) | Concept (Spine) | Concept (Neoplasia and Lesion) |
| --- | --- | --- |
| ANEURYSM | SPINE:TRAUMA | CRANIAL:NEOPLASIA |
| AVM | SPINE:TRAUMA:FRACTURE | CRANIAL:NEOPLASIA:CYST |
| CSF:LEAK | SPINE:TRAUMA:CORD | CRANIAL:NEOPLASIA:GLIOMA |
| CRANIAL:TRAUMA | SPINE:TRAUMA:DISCO-LIGAMENTOUS | CRANIAL:NEOPLASIA:MENINGIOMA |
| CRANIAL:TRAUMA:SKULL FRACTURE | SPINE:CANAL STENOSIS | CRANIAL:NEOPLASIA:METASTASIS |
| CRANIAL:TRAUMA:CONTUSIONS | SPINE:CAVERNOMA | CRANIAL:NEOPLASIA:PITUITARY |
| CRANIAL:TRAUMA:EDH | SPINE:DEGENERATIVE | CRANIAL:NEOPLASIA:SCHWANNOMA |
| CRANIAL:TRAUMA:ICH | SPINE:OTHER | CRANIAL:CAVERNOMA |
| CRANIAL:TRAUMA:IVH | OTHER:FRACTURE | SPINE:NEOPLASIA |
| CRANIAL:TRAUMA:SAH | OTHER | FISTULA |
| CRANIAL:TRAUMA:SDH | | LESION |
| CRANIAL:TRAUMA:TBI | | COMPLICATION:INFECTION |
| HYDROCEPHALUS | | |

*Table 3: Audit Categories*

The rules are continually refined by discovering additional rules based on conclusions reached after manual inspection of the processed notes - examining which words are commonly used with a given audit category and linked diagnoses. The rule refinements require confirmation by a system expert, who defined the original rules as pattern matching searches to find all notes containing an audit category. The design of the rules used here are based on the original searches, but the emphasis is on identifying all available audit categories per note, rather than returning all notes containing a category, though that is also possible.

Domain Concept identification is a third part of Concept Identification process - it identifies any remaining unidentified words by passing them through a dictionary, so that ideally all words of significance will be identified. That is, those that are useful for deducing further audit categories or can contribute to the more detailed diagnosis classification which we envisage as a final function of our application.

Having the ability to identify words and sentences individually allows us to more easily identify concepts as belonging to a word or more likely a group of words in a sentence, and thus to match individual parts of the note with appropriate concepts. We expect to be able to use these identified components as class identifiers for machine learning processing, and that subsequent analysis of outputs from machine learning processing will in turn enable further refinement of audit category rules as well as contributing to the final step of diagnosis classification. Diagnosis Classification is considered to be outside of the current method as it requires further development and integration into the current diagnosis and notes entry application, with associated considerations for adjusting the application to allow for the input of this system.

In summary, the Audit Category rule-based step finds out if a word can be identified as belonging to an audit category, and if so labels it accordingly. The Domain Concept dictionary step labels other words with information about what domain they belong to, which will provide further data to a machine learning process as well as contributing to the later analysis required for diagnosis.

After processing we have marked all words that can be identified as useful for audit classification, together with many supporting words and modifiers. Words that do not contribute to classification are likewise marked, and all other words are identified as requiring resolution. If one or more concepts are identified then it's with a high degree of certainty, and the process is easily tuneable to increase its accuracy and coverage.

# 5 EVALUATION

## 5.1 Evaluation Measures

We employ the standard evaluation metrics of Precision, Recall and F-score which clinical narrative classification typically uses to evaluate the success of an NLP system (M. H. Stanfill et al. 2010). The values for Precision, Recall and F-score are expressed as percentages.

## 5.2 Reference Standard Development

We have developed three reference standards using just the data at hand, which have unverified diagnoses only. These diagnoses were in turn divided into old and new codes, the new codes being the current codes in effect since 2008. Our domain expert had supplied audit category mappings to two thirds of the new diagnoses, so this was our working data set - consisting of 9561 records.

Our first reference standard, type A - is to accept all the 9561 records as they are, and measure our success against them.

Our second reference standard, type B - consists of diagnoses for which 50% or more of the records have terms in common found in the accompanying notes, as these are the most reliable set for training our process against. This reduced set had a total of 8544 records.

Our third reference standard, type C – is based on type A, but counts some records differently where the accompanying diagnosis is OTHER.

## 5.3 Evaluation

We evaluated the success of our audit category matching in a series of steps: Firstly, we counted as a success those where the calculated audit category exactly matched the audit category of the diagnosis – for instance both calculated and mapped codes being CRANIAL:TRAUMA:TBI.

Secondly, for those not counted in the first step we then counted those where a more generalized root portion of the calculated audit category matched to the mapped audit category – for instance the calculated audit category being CRANIAL:TRAUMA:CONTUSIONS but the mapped code being just CRANIAL:TRAUMA.

Thirdly, from the remainder we counted those where the calculated code is evidently accurate but doesn't map exactly to the mapped code, where however the code is a valid alternative code. An example is CRANIAL:TRAUMA:CONTUSIONS, which can also validly map to diagnoses having an audit category of CRANIAL:TRAUMA:TBI (a contusion of the brain is one of a number of components including haemorrhage, of a traumatic brain injury).

Fourthly we treated those where we had calculated a clearly different audit category as being unmatched, even though in many cases the codes we had calculated were a better match to the data compared with the diagnosis code mapping.

As an alternative to just categorising all of the different audit categories as unmatched, we transferred to the matched group those which clearly did have a better audit category than their accompanying diagnosis of OTHER, in order to evaluate for our third reference standard.

Finally, we had some cases where we could not find a matching audit category, often because there was insufficient information in the note, and sometimes because we have not yet developed the correct process for classifying ambiguous information.

We used the evaluation metrics of Precision, Recall, and F-Score, which are calculated as follows:
True Positive (TP):     Calculated category correctly matches the diagnosis audit category
False Positive (FP):    Calculated category doesn't match the diagnosis audit category
False Negative (FN):    No calculated category, but a valid diagnosis audit category exists
True Negative (TN):     Neither calculated or diagnosis category exists, not in consideration

Precision:     TP / (TP + FP)
Recall:        TP / (TP + FN)
F-Score:       2 x Precision x Recall/ (Precision + Recall)

Our evaluation results are summarised in Table 4.

| Reference Standard | Precision TP / (TP + FP) | | Recall (TP / TP + FN) | | F-Score |
|---|---|---|---|---|---|
| Type A | 6705 / 7910 | 84.8% | 6705 / 9561 | 70.1% | 76.8% |
| Type B | 6472 / 7199 | 89.9% | 6472 / 8544 | 75.7% | 82.2% |
| Type C | 6906 / 7910 | 87.3% | 6906 / 9561 | 72.2% | 79.0% |

*Table 4: Evaluation Results*

Clearly the results improve when we consider only those that have a good set of information in the notes, or when we take into account the likely improved accuracy of those currently matched to a diagnosis of OTHER.

One of the main goals of the system is to help discover correct audit categories in the case of those which have been inadequately coded, with those coded to category OTHER as prime candidates for recoding. As a preliminary measure of our success in processing these we can say that of the 406 records coded to OTHER, we have 210 records where we have calculated specific audit categories that are accurate when compared with their use elsewhere – and moving these into the matched group in our Type C standard reflects this. Our coding for OTHER is roughly a 50% improvement over the current situation.

Detailed analysis of the use of the OTHER category reveals that much of the time the category is used correctly, where what is being described is something that cannot be classified using the neurosurgical department's codes. They may be non-neurosurgical conditions such as cardiac problems, or they may relate to neurological symptoms such as confusion or altered conscious states that are not neurosurgical conditions – and there is no code that can be assigned to them. However, roughly half of these could be given Audit categories from areas such as Cranial Trauma and Spine, and very many of them relate to post-operative problems and should therefore be classified as Complications, and these are the codes our process has suggested.

Initial analysis of inadequate coding in other areas shows that there may be difficulty choosing codes due to the presence of valid alternatives, but that the major reason for inadequate coverage is that only one code is being used per admission. Consequently an admission with multiple problems is being coded correctly for only one problem, or an imprecise summary code may be used instead. This may be a process problem, and will be investigated as part of the conclusion of the project.

# 6   CONCLUSION

The neurosurgical audit department has specific requirements for codifying its data, which is free text expressed in an abbreviated but reasonably limited vocabulary. We decided that the best approach to this requirement was to identify what parts of the text are relevant for the generation of audit codes, by using pattern matching, some natural language processing, and dictionaries. The identification of specific terms is based on rules provided by a system expert, but enhanced by analysis of sentence structure for context and modifiers. The dictionaries help identify remaining terms and what domain they belong to, so that further processing is possible.

Our initial approach was designed to test the effectiveness of extracting information and generating audit codes just from the free text, and to measure success against the actual diagnosis attached to each record. However when extending our approach to create an application for use by the neurosurgical department we will take into account the diagnosis and other linked records, including procedures linked to the admission, and other previous diagnostic and procedural records linked to the patient.

We have developed a very efficient way to identify the majority of the useful text in the admission notes, and to generate audit codes with around 80% accuracy, as measured by F-score. The programme is very tuneable; we anticipate increasing the accuracy further with refinements to the rules, by accounting for other data, and by adding a machine learning component. The technology used is compatible with the department's systems and is capable of being integrated into both the auditing and the code entry processes.

To complete the current development of the programme we will integrate the remainder of the audit categories, and have the accuracy of the approach verified against a gold standard of records provided by the domain expert. We will seek feedback from the system expert to assist us to improve our process, and also to work out the best way of communicating our research. At the moment we are verifying our accuracy against the attached diagnoses, but we want to take account of the diagnoses going forward, so we need an external expert measure of accuracy.

We expect that the eventual application that is developed using our programme will greatly assist the neurosurgical department's auditors, by delivering the majority of the auditable data that is currently inaccessible in the notes.

## 7　REFERENCES